# High-pressure x-ray diffraction and *ab initio* study of $Ni_2Mo_3N$, $Pd_2Mo_3N$, $Pt_2Mo_3N$, $Co_3Mo_3N$, and $Fe_3Mo_3N$: Two families of ultra-incompressible bimetallic interstitial nitrides

D. Errandonea[1,*,†], Ch. Ferrer-Roca[1,†], D. Martínez-Garcia[1,†], A. Segura[1,†], O. Gomis[2,†], A. Muñoz[3,†], P. Rodríguez-Hernández[3,†], J. López-Solano[3,†], S. Alconchel[4], and F. Sapiña[5]

[1] Departamento de Física Aplicada – ICMUV, Universitat de València, Edificio de Investigación, C. Dr. Moliner 50, E-46100 Burjassot, Spain

[2] Centro de Tecnologías Físicas, Universitat Politècnica de València, E-46022 València, Spain

[3] Departamento de Física Fundamental II, Universidad de La Laguna, E-38205 La Laguna (Tenerife), Spain

[4] Departamento de Química, Facultad de Ingeniería Química, Universidad Nacional del Litoral, Santiago del Estero 2829, S3000A OM Santa Fe, Argentine

[5] Institut de Ciencia dels Materials de la Universitat de Valencia, Apartado de Correos 22085, E-46071 Valencia, Spain

**Abstract:** We have studied by means of high-pressure x-ray diffraction the structural stability of $Ni_2Mo_3N$, $Co_3Mo_3N$, and $Fe_3Mo_3N$. We also report *ab initio* computing modeling of the high-pressure properties of these compounds, $Pd_2Mo_3N$, and $Pt_2Mo_3N$. We have found that the nitrides remain stable in the ambient-pressure cubic structure at least up to 50 GPa and determined their equation of state. All of them have a bulk modulus larger than 300 GPa. Single-crystal elastic constants have been calculated in order to quantify the stiffness of the investigated nitrides. We found that they should have a Vickers hardness similar to that of cubic spinel nitrides like $\gamma$-$Si_3N_4$.

---

[*] Corresponding authors, email: daniel.errandonea@uv.es
[†] Member of the MALTA Consolider Team

## I. Introduction

A great effort is currently focused on the synthesis and characterization of superhard materials exhibiting very low compressibility and large durability [1]. The transition metal nitrides have attracted special attention for their unique properties and particularly $OsN_2$ and $IrN_2$ have been shown to have a very large bulk modulus ranking just below diamond [2, 3]. The extraordinary elastic properties of these nitrides seem to be related with the high transition-metal nitrogen coordination and the presence of strong covalent bonds. Historically, since the synthesis of cubic-$Si_3N_4$ [4], most of the attention on the search of incompressible or superhard materials has been focused on binary nitrides. However, also ternary nitrides that contain only molybdenum and other transition metal elements, due to their dense packing and the presence of short N-Mo bonds, are good candidates for being incompressible and mechanically stable materials. The advantage of these ternary nitrides against other incompressible materials is that they can be produced at ambient pressure by ammonolysis from solid precursors [5 - 9].

In this work we studied $Ni_2Mo_3N$, $Co_3Mo_3N$, and $Fe_3Mo_3N$ by means of high-pressure x-ray diffraction, using synchrotron radiation, up to 50 GPa. We found that up to this pressure the three compounds remain in their ambient-pressure cubic structure and determined their equation of state (EOS). The experimental results were complemented with *ab initio* total-energy calculations performed for $Ni_2Mo_3N$, $Pd_2Mo_3N$, $Pt_2Mo_3N$, $Co_3Mo_3N$, and $Fe_3Mo_3N$ using the density-functional theory (DFT) and the pseudopotential method. All these bimetallic nitrides have the common characteristic of having a cubic crystalline structure. Because of it, $Ni_3Mo_3N$, which has an orthorhombic crystal structure [10], was not included in the present study.

Finally, the elastic constants and associated mechanical parameters - bulk modulus (B), shear modulus (G), Young's modulus (E), and Poisson's ratio (ν) are also presented and compared with those of known superhard materials. All the studied nitrides have a bulk modulus larger than 300 GPa. Their mechanical parameters suggest that they could be suitable for applications as abrasive and cutting tools as well as scratch-resistance coatings.

## II. Experimental details

Ternary nitrides were synthesized by ammonolysis of crystalline precursors. In the Fe and Co cases, crystalline precursors were the corresponding molybdates obtained by precipitation when a 0.25 M aqueous solution of the corresponding chloride was dropwise added to an equal volume of an equimolar solution of sodium molybdate. After 24 h of refluxing, the solids were filtered off, washed with distilled water and dried at 393 K for 24 h. In the Ni case, the precursor was obtained by thermal decomposition under air (3 h, 873 K) of the product prepared by freeze drying of a 2:3 Ni:Mo aqueous solution (total cationic concentration 0.25 M). This solution was prepared by adding an aqueous solution of $NiC_4H_6O_4{\cdot}4H_2O$ to an aqueous solution of $(NH_4)_6Mo_7O_{24}{\cdot}4H_2O$ slightly acidified with acetic acid (until pH c.a. 3-4). The ammonolysis of the crystalline precursors were performed inserting into a quartz flow-through tube furnace a sample of the selected precursor (0.5 g), contained into an alumina boat. The precursor powders were heated at 5 K min$^{-1}$ to a final temperature ($T_f$) that was hold for a period of time ($t_{hold}$) under flowing ammonia (50 cm$^3$ min$^{-1}$) ($T_f$ and $t_{hold}$ were 1073 K and 12 h for Fe; 1173 K and 12 h for Co, and, 1223 K and 4 h for Ni cases). Then, the solids were cooled by quenching at room temperature (c.a. 50 K min$^{-1}$). Further details of the synthetic procedure were given elsewhere [6, 7].

Angle-dispersive x-ray diffraction (ADXRD) experiments at room temperature (RT) and high pressure with $Ni_2Mo_3N$, $Co_3Mo_3N$, and $Fe_3Mo_3N$ up to 50 GPa were carried out. Experiments were performed at beamline I15 of the Diamond Light Source using a diamond-anvil cell (DAC) and a monochromatic x-ray beam with a wavelength of 0.41328(5) Å. Samples were loaded in a 150-μm hole of an inconel gasket in a membrane-type DAC with diamond-culet sizes of 350 μm. Ruby grains were loaded with the sample for pressure determination [11]. Pressure was determined using the scale proposed by Dewaele *et al.* [12]. Either a 16:3:1 methanol-ethanol-water mixture or silicone oil was used as pressure-transmitting medium [13 – 15]. The monochromatic x-ray beam was focused down to 20 x 20 $\mu m^2$ using Kickpatrick-Baez mirrors. A pinhole placed before the sample position was used as a clean-up aperture for filtering out the tail of the x-ray beam. The images were collected using a MAR345 image plate located at 426 mm from the sample. The diffraction patterns were integrated and corrected for distortions using FIT2D. The structural analysis was performed using POWDERCELL and FULLPROF.

**III. Overview of the calculations**

The progress in the development of *ab initio* calculations enables the study of the mechanical properties of materials. The elastic constants describe these mechanical properties in the region of small deformations, where the stress-strain relations are still linear. In this work we are dealing with cubic crystals, and for cubic symmetry there are three independent elastic constants, $C_{11}$, $C_{12}$, and $C_{44}$ [16]. The macroscopic stress in the solid can be computed for a small strain by the use of the stress theorem [17]. Alternatively, it can be also calculated using density-functional perturbation theory (DFPT) [18]. The small total-energy differences between the different strained states should be calculated with a high precision.

All of the calculations reported here have been performed using DFT and DFPT, with the local-density approximation (LDA) prescription for the exchange-correlation energy [19] implemented in the Vienna *ab inito* simulation package (VASP) [20, 21] with the pseudopotential method. The set of plane waves employed extended up to a kinetic energy cutoff of 540 eV for all the studied nitrides. Such a large cutoff was required to achieve highly converged results within the projector-augmented wave (PAW) scheme [22, 23] with the N atom. The PAW method takes into account the full nodal character of all electron charge density distribution in the core region. We used a dense k-grid for Brillouin zone (BZ) integrations in order to assure highly converged results to about 1-2 meV per formula unit. We also followed an accurate procedure in the calculations in order to obtain very well converged forces which were used for the relaxation at different volumes. The structures were fully relaxed to their equilibrium configuration through the calculation of the forces on atoms and the stress tensor. In the relaxed equilibrium configuration, the forces are less than 0.004 eV/Å and the deviation of the stress tensor from a diagonal hydrostatic form is less than 0.1-0.2 GPa. From the energy-volume (E-V) curves obtained from the *ab initio* calculations the equilibrium volume, bulk modulus and pressure derivatives are obtained using a fourth-order Birch-Murnaghan EOS [24].

The bulk modulus can be also obtained from the elastic constants. For a cubic crystal it can be expressed in the form [16]:

$$B = \frac{c_{11} + 2c_{12}}{3} \quad (1)$$

This parameter is the inverse of the compressibility and measures the resistance of the material to a uniform hydrostatic pressure. We also report the isotropic shear modulus, G, elastic moduli, E, and the Poisson's ratio, υ. These parameters describe the major elasticity properties for a material. In our case they are defined by the following equations [16], where for the shear modulus we assumed the average of the Voigt and Reuss bounds [25].

$$G=\frac{1}{2}\left[\frac{c_{11}-c_{12}+3c_{44}}{5}+\frac{5c_{44}(c_{11}-c_{12})}{4c_{44}+3(c_{11}-c_{12})}\right]; \tag{2}$$

$$E=\frac{9BG}{3B+G}; \tag{3}$$

and

$$\upsilon=\frac{E-2G}{2G}. \tag{4}$$

## IV. Results and discussion

### A. High-pressure ADXRD studies

At ambient pressure (0.0001 GPa), the obtained diffraction pattern for $Ni_2Mo_3N$ corresponded to the previously reported cubic structure (space group: *P4₁32,* No. 213, Z = 4, *a* = 6.634(4) Å) [5 - 7], with no indication of any additional phase in it. The structure of $Ni_2Mo_3N$ is represented in Fig. 1. This structure consists of a β-Mn arrangement of Ni (Pd or Pt) atoms, with N atoms occupying distorted octahedral sites [6 - 8]. The Ni atoms are located in a 12-fold pseudo-icosahedral coordination surrounded by three Ni and nine Mo atoms. The Mo atoms are located in a 14-fold pseudo-tetradecahedron site surrounded by six Ni, six Mo, and two N. Finally, each N is surrounded by six Mo atoms, defining an octahedron with short Mo-N bond distances (2.0794 Å), these octahedra share corners. Fig. 2 shows ADXRD data for $Ni_2Mo_3N$ at several selected pressures. Under compression, the only changes we observed in the x-ray diffraction patterns are a slight shift of Bragg peaks towards high 2θ angles and the typical peak broadening of DAC experiments which in our case is detectable around 40 GPa. [26 - 28]. No evidence of pressure-induced phase transition or chemical decomposition of $Ni_2Mo_3N$ is detected up to 50 GPa and the ambient-pressure diffraction pattern is fully recovered upon decompression. The possibility of a pressure-induced tetragonal distortion of the cubic structure is also excluded. We also verified that the use of different pressure media does not affect the experimental results. All the measured x-ray diffraction patterns can be indexed within the $P4_132$ cubic structure. From the ADXRD data, we obtained the evolution with pressure of the lattice parameters (and unit-cell volume) of $Ni_2Mo_3N$. We also refined the internal coordinates of the Ni and Mo atoms (the only two free coordinates in the structure). We found that, within the pressure range of the experiments, the pressure change of the *x* coordinate of Ni (Wyckoff position 8c) and the *y* coordinate of Mo (Wyckoff position 12d) is smaller than the experimental uncertainty. The mean values of these coordinates are $x_{Ni}$=0.189(4) and $y_{Mo}$=0.047(1). Therefore, we concluded that the pressure effect on the atomic positions can be neglected. The pressure evolution of the unit-cell volume of $Ni_2Mo_3N$ is plotted in Fig. 3. There it can be seen that this nitride is highly incompressible. In a 50 GPa range, the volume change induced is only 10%. The present pressure-volume data have been analyzed using a third-order Birch–Murnaghan EOS [24]. The obtained volume at ambient pressure ($V_0$), bulk modulus ($B_0$), and its pressure derivative ($B_0$') are summarized in Table I. The implied value of the second pressure derivative of the bulk modulus in a third-order truncation of the Birch-Murnaghan EOS is also given: $B_0'' = -[(B_0'-4)(B_0'-3)+35/9]/B_0$. The bulk modulus of $Ni_2Mo_3N$ (330 GPa) is larger than that of cubic-spinel nitrides (e.g. $\gamma$-$Si_3N_4$) [1], comparable to that of many binary transition-metal nitrides (e.g. PtN) [3], and only 10% smaller than that of cubic BN [29].

In the case of $Co_3Mo_3N$, and $Fe_3Mo_3N$ we observed a similar behavior when compared to $Ni_2Mo_3N$. At ambient conditions, the obtained diffraction patterns corresponded to the known cubic structure (space group: $Fd\bar{3}m$, No. 227, Z = 16, *a* =

11.027(9) Å and $a$ = 11.086(94) Å, respectively) [6, 30], with no indication of any additional phase in it. The structure of $Fe_3Mo_3N$ is represented in Fig. 4. This structure is similar to the η-carbide structure and can be described in terms of two interpenetrating networks of corner shared $NMo_6$ octahedra and $Fe[Mo_6Fe_6]$ and $Fe[Mo_6Fe_4N_2]$ pseudo-icosahedra [6, 30]. Both compounds have short Mo-N bond distances like $Ni_2Mo_3N$ (2.1128 Å and 2.1137 Å). Under compression, we observed that in both compounds all the Bragg peaks shift smoothly with pressure and that all of them could be assigned to the $Fd\overline{3}m$ structure up to 50 GPa. No evidence of pressure-induced phase transition, chemical decomposition, or structural distortion is detected. Ambient-pressure diffraction patterns are fully recovered after decompression. From the x-ray diffraction data, we obtained the evolution with pressure of the lattice parameters (and unit-cell volume). We also refined the atomic positions of the Co (Fe) and Mo atoms that have free coordinates. In $Co_3Mo_3N$ and $Fe_3Mo_3N$, the resulting coordinates were: Co and Fe (Wyckoff position 32e) $x_{Co}$ = 0.292(4) and $x_{Fe}$ = 0.294(4), respectively; Mo (Wyckoff position 48f) $x_{Mo}$ = 0.323(5) and $x_{Mo}$ = 0.321(5), respectively. These positions agree with those reported in the literature [6, 30] and the effect of pressure on them is comparable with the uncertainty of the experiments. The pressure dependences of the volume of both nitrides are plotted in Fig. 3. The pressure-volume data were analyzed using a third-order Birch–Murnaghan EOS [24]. The EOS parameters are summarized in Table I. Again both compounds are highly incompressible. The obtained bulk muduli are: 350 GPa ($Co_3Mo_3N$) and 368 GPa ($Fe_3Mo_3N$). Thus, these compounds turn out to be even less compressible than $Ni_2Mo_3N$. The volume decrease for both nitrides from ambient pressure to 50 GPa is smaller than 9%. The bulk modulus of $Fe_3Mo_3N$ turns out to be only 1% smaller than the one of cubic BN.

B. ***Ab initio* Results**

We compare now the experimental data presented with the results obtained from our total-energy calculations. Fig. 5 shows the energy-volume curves for the different nitrides considered, from which the structural parameters and EOS can be extracted. The cubic structure with space group $P4_132$ is considered for $Ni_2Mo_3N$, $Pd_2Mo_3N$, and $Pt_2Mo_3N$ and the cubic structure with space group $Fd\overline{3}m$ for $Co_3Mo_3N$ and $Fe_3Mo_3N$. The obtained lattice parameter at ambient pressure is: 6.5601 Å for $Ni_2Mo_3N$, 6.7682 Å for $Pd_2Mo_3N$, 6.8038 Å for $Pt_2Mo_3N$, 10.8158 Å for $Co_3Mo_3N$, and 10.8401 Å for $Fe_3Mo_3N$. These values compare well with the experimental results, with differences within the typical reported systematic errors in DFT calculations [31, 32]. A similar degree of agreement exists for the calculated values of the internal parameters. Calculations also confirm that there is no important effect of the pressure on the atomic positions of the atoms in all the nitrides. In addition, a fit to the results shown in Fig. 5 with a Birch-Murnaghan fourth-order EOS [24] gives values in good agreement with experimental results, as shown in Table I. The equilibrium volume $V_0$ is underestimated by the calculations by 1-2% in $Ni_2Mo_3N$, $Pd_2Mo_3N$, and $Pt_2Mo_3N$ and by 5-7 % in $Co_3Mo_3N$ and $Fe_3Mo_3N$. In turn the bulk modulus $B_0$ is underestimated by a similar proportion but the relative differences between the experimental bulk moduli of the three compounds are correctly predicted. The high bulk modulus calculated for the five compounds supports the experimental conclusions indicating the studied nitrides are ultra-incompressible compounds.

Let us concentrate now on the calculation of the elastic constants. The computed elastic constants tensor is shown in Table II. The elastic constants are compared with those of known superhard materials [1, 29, 33, 34, 35]. The whole set of elastic constants calculated fulfill the stability criteria for a cubic crystal:

$(c_{11} - c_{12}) > 0$, $c_{11} > 0$, $c_{44} > 0$, and $(c_{11} + 2c_{12}) > 0$, pointing to the mechanical stability of the five nitrides. It is also worth to know that the bulk modulus computed from the values of the elastic constants is in good agreement with the experiments and the one obtained from the total-energy calculations, giving therefore a consistent estimation of the compressibility of bimetallic interstitial nitrides. In order to discuss elasticity in detail G was also calculated (see Table II). The obtained values are comparable with the shear modulus of cubic-SiC (carborundum) [33, 35, 36] and cubic spinel-type $Hf_3N_4$ [35]. It is also important to notice that in the five nitrides B/G > 2. The critical B/G value for ductile and brittle materials is 1.75, indicating that the studied compounds are ductile materials. Additionally, we calculated the Zener anisotropy ratio defined as $A=2C_{44}/(C_{11} - C_{12})$ which is an important physical quantity related to structural stability. It gives the deviation from elastic isotropy in cubic crystals. This quantity is calculated to be 1.57, 1.33, 1.17, 1.09, and 1.16 for the Ni, Pd, Pt, Co, and Fe nitrides respectively. The more this parameter differs from 1, the more elastically anisotropic the crystalline structure. In our case, the five compounds show a small elastic anisotropy similar to that of extremely stable metals. Indeed in a bcc metal like tungsten (W) the Zener anisotropy is close to 1 and in fcc metals like aluminum (Al) close to 1.6. Our values are in between these two limits. The cubic metals W and Al (and similar ones like Mo, Ta, Au, or Pt) are quite stable upon compression; they do not undergo phase transitions even at Mbar pressures. Then from the elastic point of view we expect also the studied nitrides to remain stable up to extreme pressures; in agreement with our experimental findings.

Two other numbers are important for technological and engineering applications: the Young's modulus and Poisson's ratio. The first one is defined as the ratio between stress and strain and is used to provide a measure of the stiffness of a solid. The second one quantifies the stability of the crystal against shear. The obtained values for E range from 332 to 392 GPa. These values are comparable to the Young's modulus of cubic $Hf_3N_4$ [33] and BN [29]. They indicate that the five nitrides are stiff materials. Regarding Poisson's ratio, all calculated values are close to 0.3 a value similar to the Poisson's ratio in steel and cubic BN. The obtained values for ν mean that the studied materials are substances with predominantly central internal forces. Note that the Poisson's ratio depends monotonically upon G/B. While the bulk modulus is more sensitive to the spatially averaged electron density, the shear modulus is a quantity that accounts for the non-uniform distribution of the electron density. In the studied nitrides the bulk modulus is always larger than the double of the shear modulus leading to Poisson's ratios close to 0.3. This is characteristic of their stiff lattice, a direct consequence of the fact that the Mo atoms are coordinated by a high number of nitrogen atoms. Such coordination tends to bring the system towards a quasi-spherical disposition of nearest neighbors with the consequent loss in the bond bending forces [29], causing the central forces to dominate the mechanical properties. According to our results, this behavior is not modified upon compression.

For completeness we have also calculated the pressure dependence of the elastic constants in the five studied nitrides. The obtained results are summarized in Fig. 6. In the pressure range covered by our calculations, the elastic constants and their relations satisfy all off the stability criteria. In particular not only the elastic constants, but also $C_{11} - C_{12}$, increase under compression. In Table III we summarize the values obtained at 45 GPa from the elastic constants for B, G, E, and ν. The three first magnitudes increase more than 30% from ambient pressure to 45 GPa. However, the Poisson ratio slightly changes in all the compounds staying close to 0.33; i.e the properties of the studied nitrides are still dominated by central forces under extreme compression. Regarding the effects of pressure on the ductility of binary nitrides, we found that B/G increases under pressure reaching at 45 GPa for the different compounds values that ranges from 2.20 to 2.68 GPa. This fact indicates that under compression the ductility of binary interstitial nitrides is enhanced. The effects of pressure on the Zener ratio can be also seen in Table III. The calculations indicate that in all the studied compounds this parameter increases under compression. However, even at pressures of 45 GPa the Zener ratio still takes similar values than in extremely stable cubic metals, retaining a weak elastic anisotropy under compression. This is consistent with the fact that the same compounds remain in the ambient pressure phase even at 50 GPa.

### C. Hardness and bond compressibility

Let us now comment here on the potential hardness of the bimetallic interstitial nitrides. It is acknowledged that the bulk modulus or shear modulus can be a measure of the hardness in an indirect way [37]. Materials with high B and G are likely to be hard materials. Employing the correlation between the shear modulus and the Vickers hardness reported by Teter [37] for a wide variety of hard materials, it is possible to provide an estimation of the indentation hardness of interstitial nitrides. By using the shear modulus reported in Table II we found a theoretical Vickers hardness of 20.5 GPa for $Ni_2Mo_3N$, 17.1 GPa for $Pd_2Mo_3N$, 19.4 GPa for $Pt_2Mo_3N$, 18.3 GPa for $Co_3Mo_3N$, and 20.3 GPa for $Fe_3Mo_3N$. These values are close to the Vicker hardness of $\gamma$-$Si_3N_4$ (21±3 GPa) [1], thus pointing to a very similar resistance to plastic deformations. They are however only 18% that of diamond (96 ±3 GPa) [29], which remains by far the hardest known material.

Finally, we would like to relate the incompressible and non-deformable structure of bimetallic interstitial nitrides with their common Mo-N bonds. The five studied compounds have a characteristic short covalent bond between Mo and N atoms. This bond distance in the five nitrides is pretty similar to the Mo-N distance in rock-salt-type molybdenum nitride, i.e. 2.095 Å [38]. In our case the Mo-N bonds form a very strong three-dimensional covalent network, which gives the studied materials their particular mechanical properties. The Mo-N bond is quite uncompressible - e.g. it decreases in $Ni_2Mo_3N$ from 2.0797 Å at ambient pressure to 2.0002 Å at 50 GPa – making consequently the studied nitrides also uncompressible.

## V. Concluding Remarks

We studied experimentally and theoretically the structural and mechanical properties of bimetallic interstitial nitrides. We found that these compounds are stable in their cubic ambient-pressure structure at least up to 50 GPa. The five studied compounds have a large bulk modulus. The rest of the mechanical parameters are also calculated suggesting together with the bulk modulus that interstitial nitrides are incompressible and stiff. In addition, an estimation of the hardness indicates that these compounds can be as hard as superhard spinel-type nitrides. However, they have the advantage that can be synthesized in large amounts without the use of high-pressure or high-temperature conditions. We hope that this work will stimulate further research on this technological and interesting family of nitrides.

## Acknowledgments

We acknowledge the financial support of the Spanish MCYT (Grants MAT2007-65990-C03-01/03, MAT2010-21270-C04-01/03, CSD2007-00045 and MAT2009-14144-CO3-03). X-ray diffraction experiments carried out with the support of the Diamond Light Source at the I15 beamline, proposal EE3652. The authors thank A. Kleppe for technical support during the experiments.

**Table I:** Experimentally determined and calculated EOS parameters.

| Compound | | $V_0$ (Å$^3$) | $B_0$ (GPa) | $B_0$' | $B_0$'' (GPa$^{-1}$) |
|---|---|---|---|---|---|
| $Ni_2Mo_3N$ | Experiment | 291.96(2) | 330(8) | 4.5(5) | -0.014[*] |
| | Theory | 282.131 | 317.53 | 4.54 | -0.016 |
| $Pd_2Mo_3N$ | Experiment | 316.89[a] | | | |
| | Theory | 310.046 | 312.11 | 4.94 | -0.028 |
| $Pt_2Mo_3N$ | Experiment | 319.43[b] | | | |
| | Theory | 314.966 | 341.80 | 4.81 | -0.020 |
| $Co_3Mo_3N$ | Experiment | 1340.8(6) | 350(8) | 4.7(6) | -0.014[*] |
| | Theory | 1265.24 | 333.26 | 4.74 | -0.028 |
| $Fe_3Mo_3N$ | Experiment | 1362.4(6) | 368(9) | 4.2(4) | -0.011[*] |
| | Theory | 1273.79 | 351.35 | 3.62 | -0.0006 |

[*] Implied value. [a] Ref. [8]. [b] Ref. [9].

**Table II:** Calculated zero-pressure elastic constants of the studied nitrides. The elastic moduli (B, G, E) and Poisson's ratio (ν) have been calculated using eqs. 1 – 4. The different parameters are compared with those of know superhard materials.

| Compound | $c_{11}$ (GPa) | $c_{12}$ (GPa) | $c_{44}$ (GPa) | B (GPa) | G (GPa) | E (GPa) | ν |
|---|---|---|---|---|---|---|---|
| $Ni_2Mo_3N$ | 471.45 | 240.42 | 181.60 | 317.43 | 151.47 | 392.06 | 0.2941 |
| $Pd_2Mo_3N$ | 442.36 | 229.73 | 141.41 | 300.60 | 126.14 | 331.99 | 0.3159 |
| $Pt_2Mo_3N$ | 519.91 | 258.81 | 152.43 | 345.84 | 143.26 | 377.64 | 0.3180 |
| $Co_3Mo_3N$ | 503.65 | 247.95 | 138.96 | 333.18 | 134.40 | 355.42 | 0.3222 |
| $Fe_3Mo_3N$ | 524.27 | 250.08 | 159.13 | 341.47 | 149.91 | 392.33 | 0.3085 |
| cubic-$Hf_3N_4$[a] | 493.2 | 167.3 | 152.4 | 275.9 | 156.6 | 395.0 | 0.2610 |
| γ-$Si_3N_4$[b] | 551-500 | 191-159 | 349-327 | 317-300 | 267-251 | 623-577 | 0.165-0.147 |
| Cubic-SiC[c] | 390 | 142 | 256 | 225 | 192 | 448 | 0.1680 |
| cubic-BN[d,e] | 824.64 | 182.58 | 495.80 | 369-382 | 973-840 | 409 | 0.3765 |
| Diamond[d,e] | 1116.6 | 137.22 | 604.31 | 442-433 | 1162 | 534-544 | 0.1760 |

[e] Ref. [35], [b] Ref. [1], [c] Ref. [33], [d] Ref. [29], [e] Ref. [38].

**Table III:** Calculated elastic moduli (B, G, E) and Poisson's ratio (ν) at 45 GPa

| Compound | B (GPa) | G (GPa) | E (GPa) | ν | Zener ratio |
|---|---|---|---|---|---|
| $Ni_2Mo_3N$ | 446 | 202 | 526 | 0.3020 | 1.82 |
| $Pd_2Mo_3N$ | 472 | 183 | 486 | 0.3279 | 1.58 |
| $Pt_2Mo_3N$ | 511 | 210 | 554 | 0.3190 | 1.29 |
| $Co_3Mo_3N$ | 488 | 196 | 518 | 0.3214 | 1.28 |
| $Fe_3Mo_3N$ | 485 | 181 | 483 | 0.3342 | 1.23 |

**Figure Captions**

**Figure 1:** Schematic representation of the structure of $Ni_2Mo_3N$. The unit-cell and atomic bonds are shown. Large black circles: Mo, small black circles: N, and grey circles: Ni.

**Figure 2:** Selection of room-temperature ADXRD data of $Ni_2Mo_3N$ at different pressures (indicated in the plot). Experiments performed using methanol-ethanol-water as pressure transmitting medium. In all diagrams, the background was subtracted.

**Figure 3:** Unit-cell volume of $Ni_2Mo_3N$, $Co_3Mo_3N$, and $Fe_3Mo_3N$ under pressure. The symbols represent the present experimental data. Empty symbols: experiments performed under methanol-ethanol-water. Solid symbols: experiment performed under silicone oil. The solid lines are the reported EOS. For $Co_3Mo_3N$ and $Fe_3Mo_3N$ we plotted V/4 to facilitate the comparison with $Ni_2Mo_3N$.

**Figure 4:** Schematic representation of the structure of $Fe_3Mo_3N$. The unit-cell and atomic bonds are shown. Large black circles: Mo, small black circles: N, and grey circles: Fe.

**Figure 5:** Total-energy versus volume from *ab initio* calculations for the studied nitrides.

**Figure 6:** Calculated pressure dependence of the elastic constants for different nitrides.

**Figure 1**

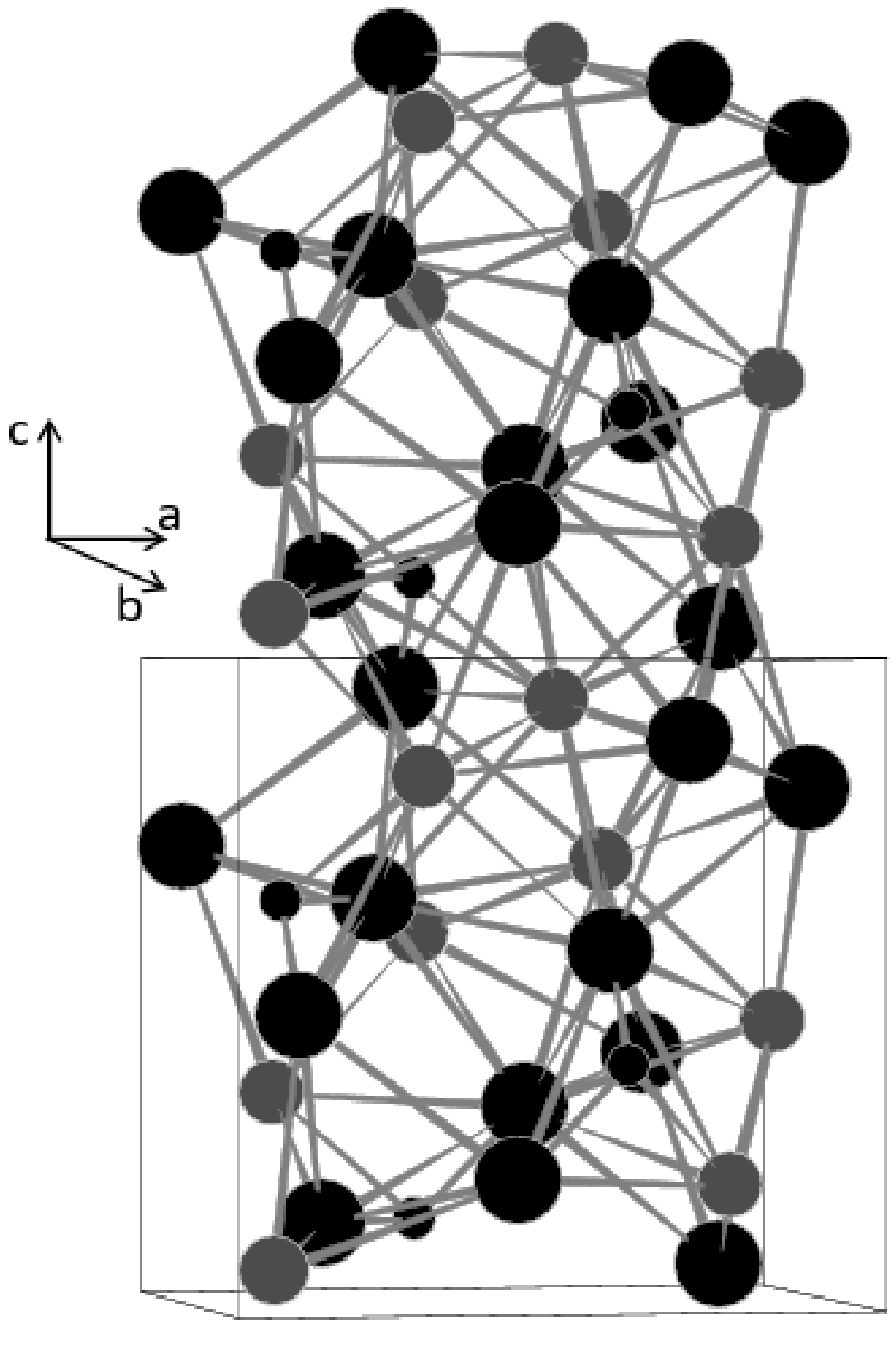

**Figure 2**

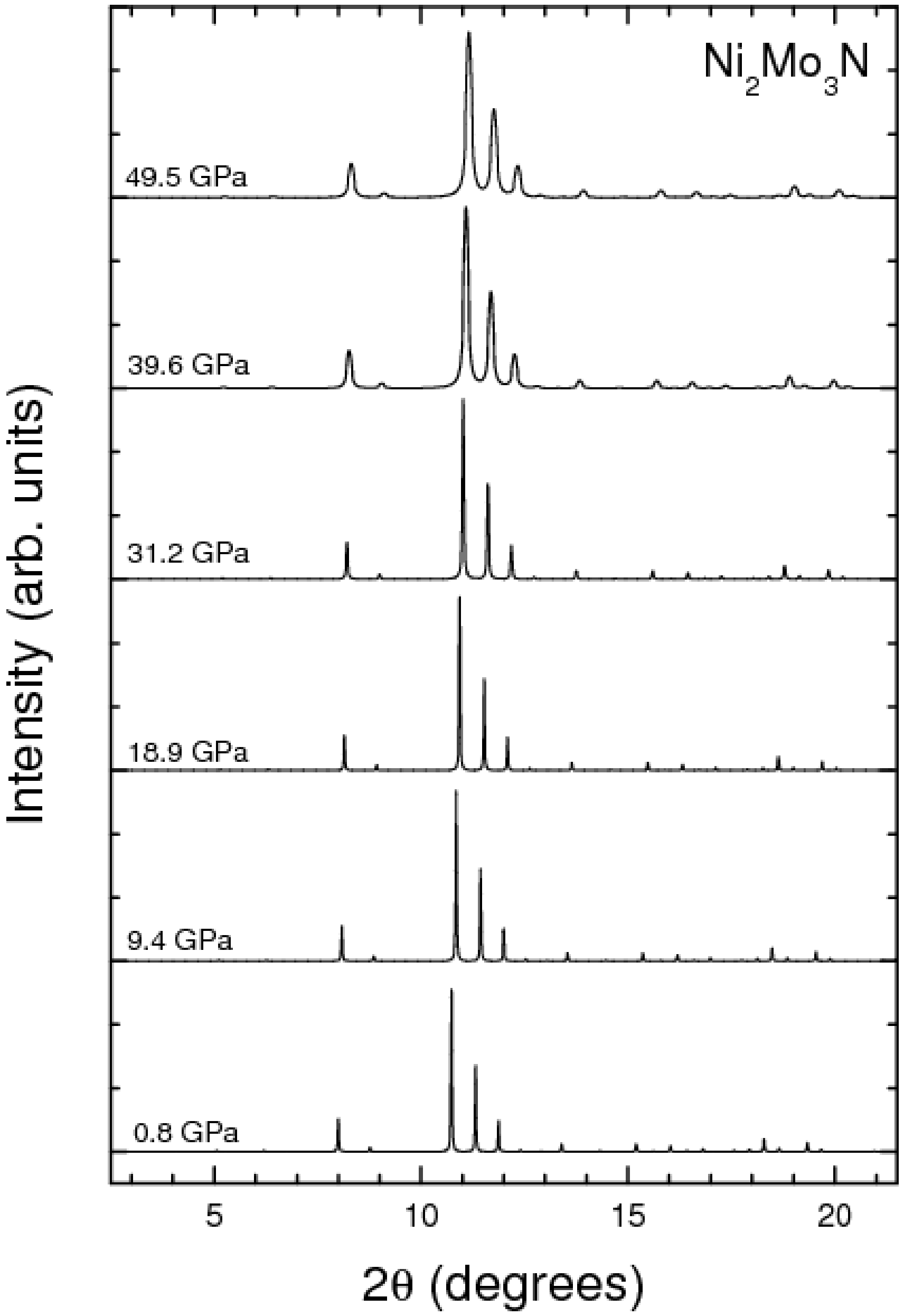

**Figure 3**

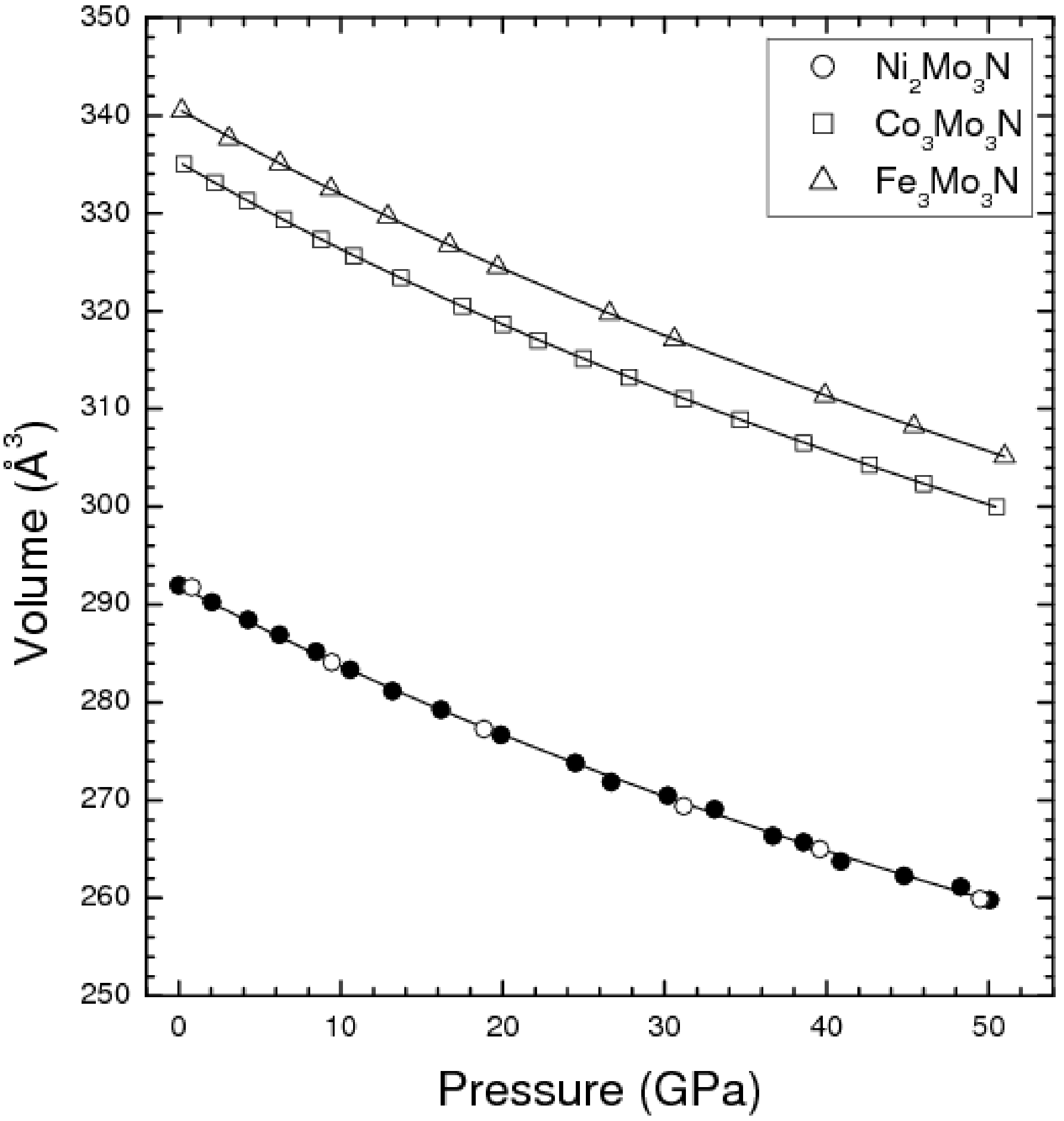

**Figure 4**

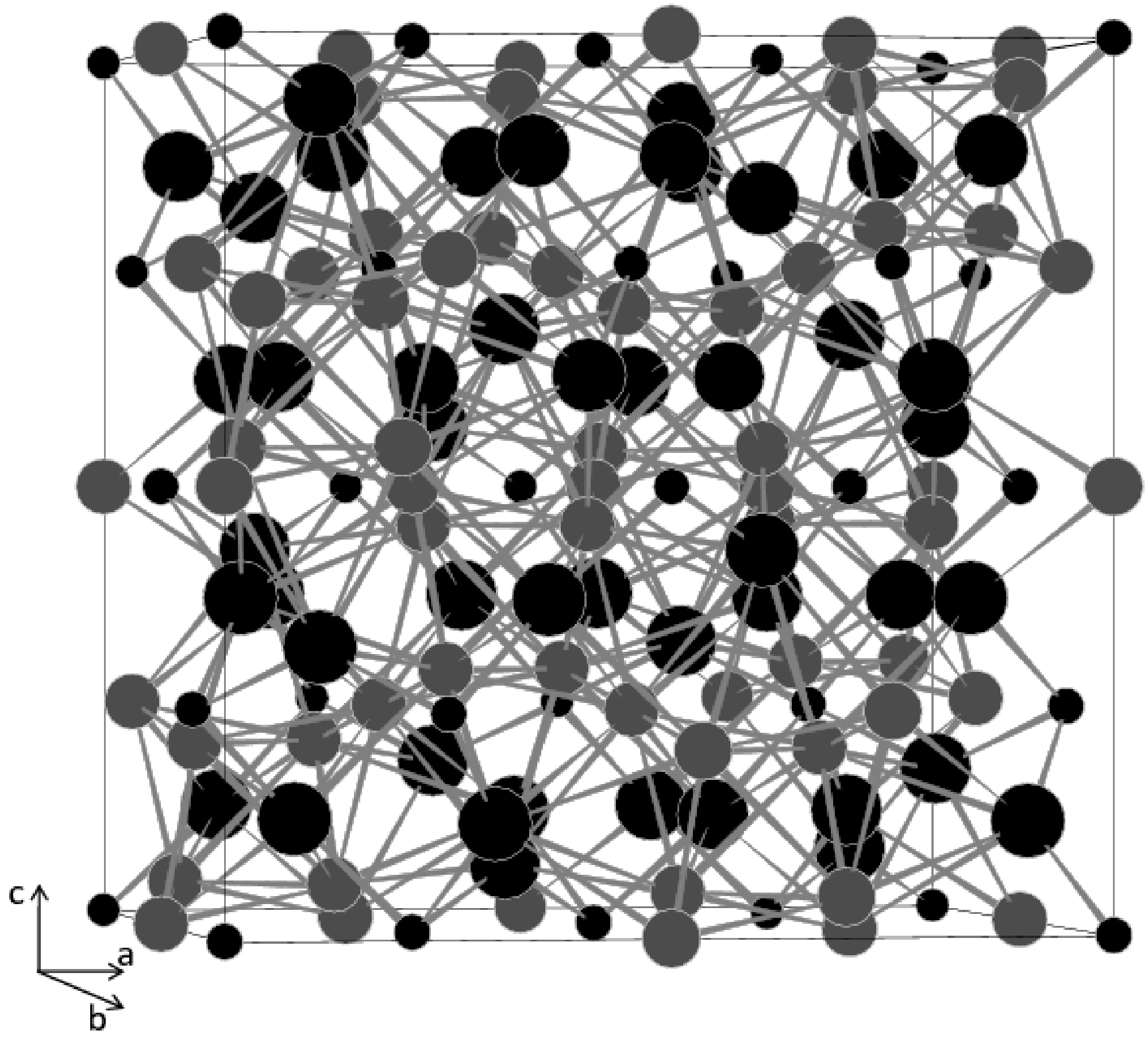

**Figure 5**

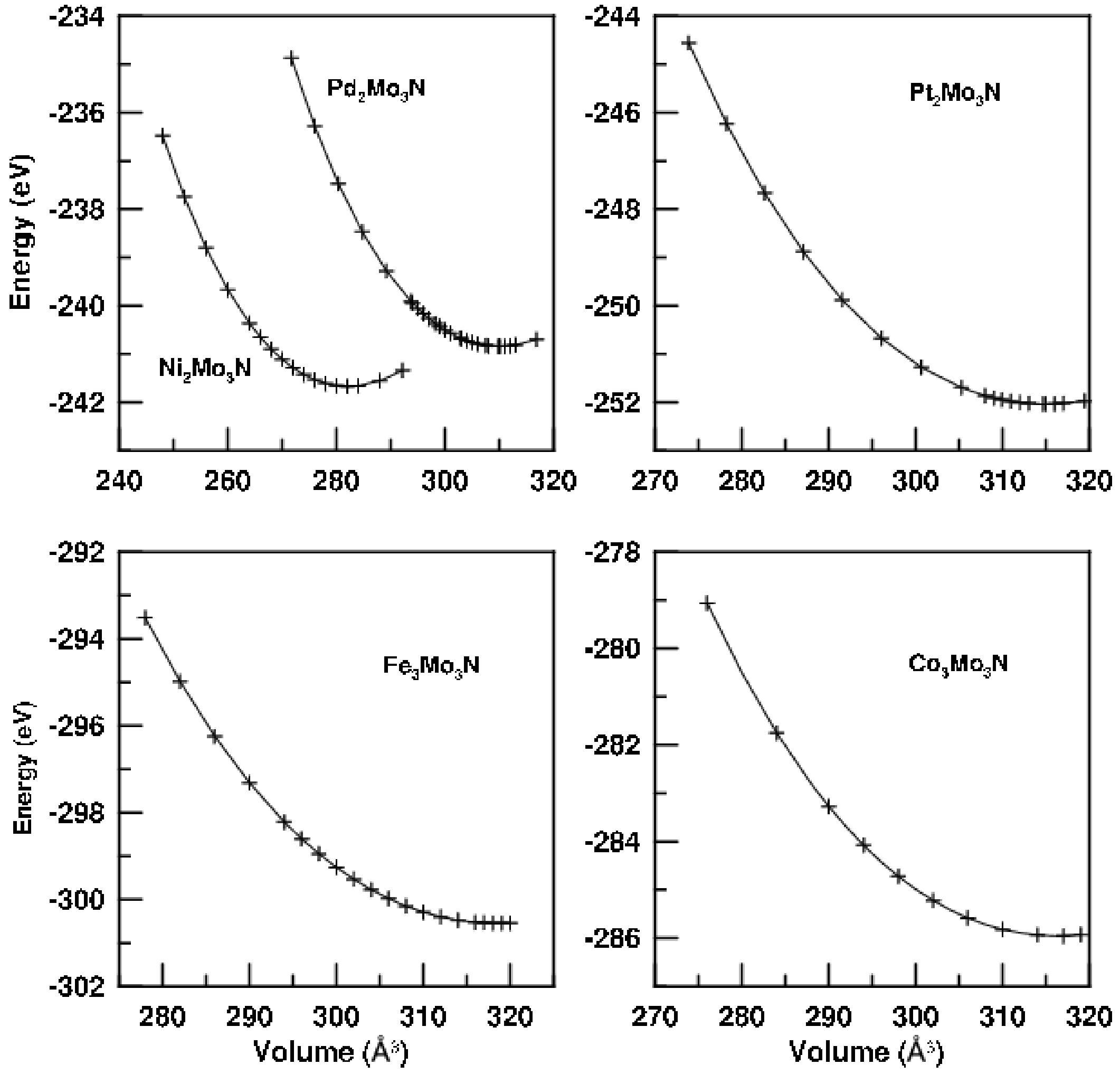

**Figure 6**

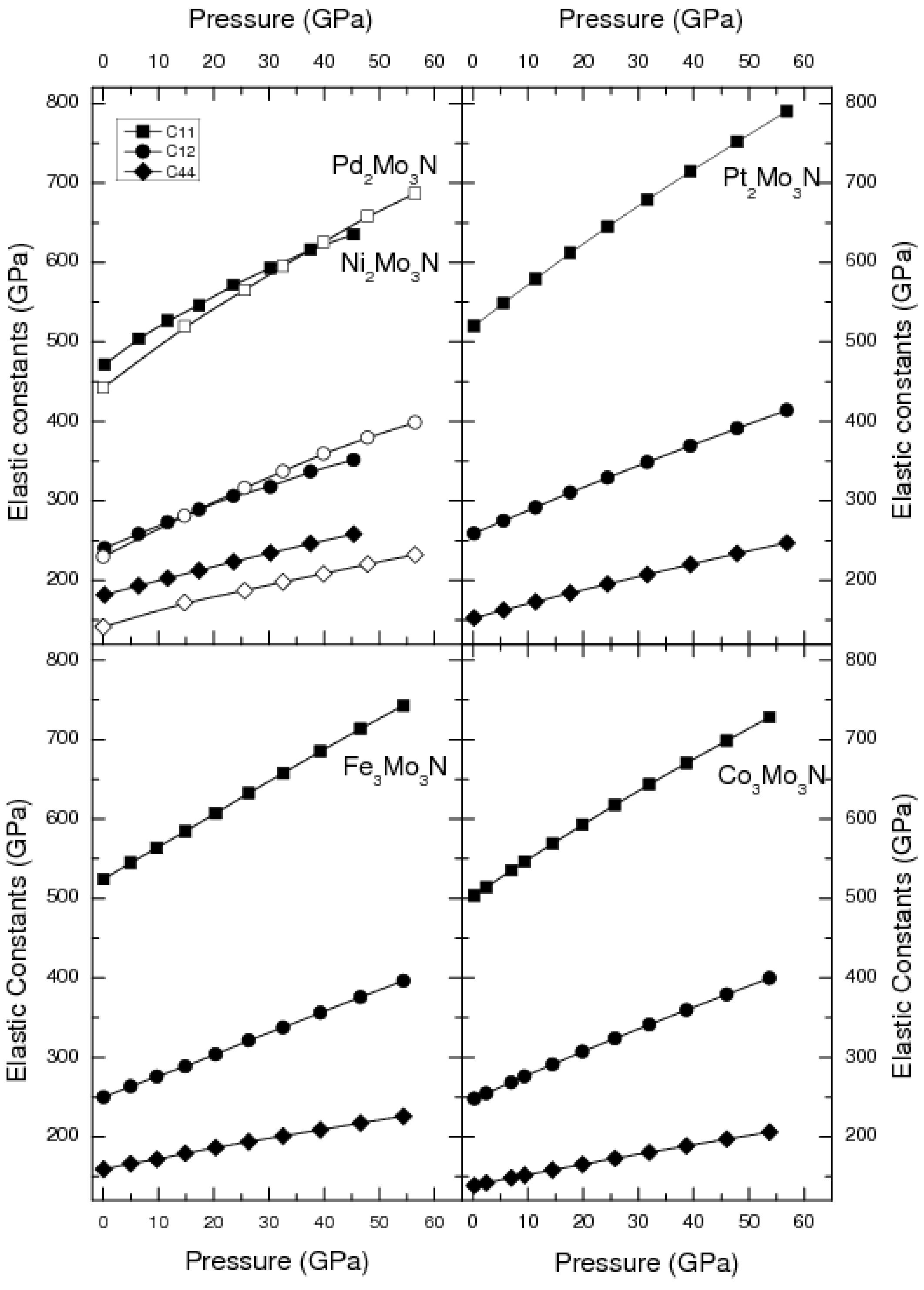